%% file: main.tex
\documentclass[letterpaper, 10pt, conference]{ieeeconf}  
\usepackage[T1]{fontenc}
\usepackage{CJKutf8}
\usepackage[english]{babel}
\usepackage{tipa}
\usepackage{multirow}
\usepackage{setspace}
\usepackage{comment}
\usepackage{cite}
\usepackage{graphicx}
\usepackage{booktabs}
\usepackage{url}
\usepackage{subfigure}
\usepackage{boldline, multirow, multicol}
\usepackage{subfig}
\usepackage{pgfplots, mathtools}
\usepackage{xcolor}
\usepackage{subcaption}
\usepackage{stackengine}

\IEEEoverridecommandlockouts                              

\newcommand{\shortname}{C2C}

\title{\LARGE \bf
C2C: Cough to COVID-19 Detection in BHI 2023 Data Challenge}
\author{Woo-Jin Chung, Miseul Kim, and Hong-Goo Kang 
\thanks{All authors are with Digital Signal Processing \& Artificial Intelligence Lab, School of Electrical and Electronic Engineering, Yonsei University, Seoul, South Korea {\tt\small \{woojinchung\}@dsp.yonsei.ac.kr}}
}

\begin{document}

\maketitle
\thispagestyle{empty}
\pagestyle{empty}

\input{0_Abstract}

\input{1_Introduction}

\input{2_Preprocessing}

\input{3_Proposed_model}

\input{4_Experiments}

\input{5_Conclusions}

\bibliographystyle{IEEEtran}
\bibliography{my_ref}

\end{document}

%% file: 0_Abstract.tex
\begin{abstract}
This report describes our submission to BHI 2023 Data Competition: Sensor challenge.
Our Audio Alchemists team designed an acoustic-based COVID-19 diagnosis system, Cough to COVID-19 (C2C), and won the 1st place in the challenge.
C2C involves three key contributions: pre-processing of input signals, cough-related representation extraction leveraging Wav2vec2.0, and data augmentation.
Through experimental findings, we demonstrate C2C's promising potential to enhance the diagnostic accuracy of COVID-19 via cough signals.
Our proposed model achieves a ROC-AUC value of 0.7810 in the context of COVID-19 diagnosis.
The implementation details and the python code can be found in the following link: https://github.com/Woo-jin-Chung/BHI\_2023\_challenge\_Audio\_Alchemists

\end{abstract}

\vspace{-3pt}

%% file: 1_Introduction.tex
\section{Introduction}
\label{lab:intro}
\vspace{-2pt}

The BHI 2023 Sensor Informatics Challenge centers on constructing acoustic-based COVID-19 diagnosis systems. 
These systems classify sound signals (coughs and deep breaths) into two categories: positive or negative for COVID-19. 
Our focus is primarily on utilizing cough recordings, resulting in the development of C2C, a novel COVID-19 diagnosis model that combines a pre-trained Wav2vec 2.0~\cite{baevski2020wav2vec} and ECAPA-TDNN~\cite{desplanques20_interspeech} neural network architectures. 
We significantly improve the accuracy of classification by employing an efficient network architecture, optimizing data pre-processing, and incorporating data augmentation during network training.

\vspace{-2pt}

%% file: 2_Preprocessing.tex
\section{Pre-processing}
\label{sec:pre-processing}
\vspace{-2pt}

To ensure stable network training, we focus on detecting the time steps of cough signals, excluding unnecessary silent periods and noise segments occurring in between coughing sections.
In particular, we employ short-time energy (STE) analysis with a 22.5 ms window length (and an 11.25 ms hop length) to identify the start and end points of coughing sounds.
First, we normalize the signals by their maximum amplitude to manage the dynamic amplitude fluctuations. 
Secondly, the start time steps of coughing are determined by detecting STE values that surpass a specific threshold, which is set at 14.5.
Third, we pinpoint the coughing region by identifying the nearest positions to the subsequent cough starting frame where the STE value exceeds 0.1.
Finally, by using the extracted time steps of cough signals, we effectively segment the frames activated by coughs from the original signals.
The detailed implementation configuration can be found in the supported GitHub repository.

%% file: 3_Proposed_model.tex
\section{Proposed model}
\label{sec:proposed_model}
\vspace{-2pt}
C2C is developed with the purpose of diagnosing COVID-19 using cough signals as a diagnostic tool.
Fig. 1-(a) illustrates the structure of the proposed model, which comprises three modules: a feature extractor, an E-T encoder, and a classifier.
The pre-trained feature extractor obtains rich cough-related representations from the input signals. 
The E-T encoder produces COVID-19 diagnosis features using the representations obtained from the feature extractor.
Finally, the classifier then generates probabilities that indicate the COVID-19 status of the patient.

\noindent \textbf{Feature extractor.}
Wav2vec 2.0~\cite{baevski2020wav2vec}, a self-supervised learning framework, demonstrates exceptional proficiency in extracting speech representations from unlabeled audio signals.
By combining CNN and Transformer layers, the model effectively captures abundant feature information from raw speech waveforms\footnote{https://github.com/facebookresearch/fairseq/tree/ust/examples/wav2vec}.
We utilize Wav2vec 2.0 as the feature extractor to enhance our cough-based COVID-19 diagnosis system.
Moreover, we fine-tune all layers of the Wav2vec 2.0 model using cough recordings to extract features specifically related to cough signals.

\noindent \textbf{E-T encoder.}
ECAPA-TDNN, a model originally designed for speaker verification, leverages the capabilities of the Convolutional Neural Network (CNN) architecture in the context of speaker verification tasks.
It employs multi-layer feature aggregation and self-attentive pooling to generate a unified feature representation.
Leveraging the capabilities of the ECAPA-TDNN model, our E-T block follows the backbone architecture of ECAPA-TDNN while reducing the hidden channel sizes by one eighth.

\noindent \textbf{Classifier.}
After removing the final few fully connected layers from the vanilla ECAPA-TDNN, we add two fully connected layers to predict the probabilities of the COVID status.
The use of a sigmoid activation function restricts the range of output features to be within the 0 to 1 range.

\noindent \textbf{Training criteria.}
The model is trained in an end-to-end fashion by minimizing the binary cross-entropy loss between the target COVID status and the predicted probabilities.
\vspace{-3pt}

%% file: 4_Experiments.tex
\section{Experiments}
\label{sec:experiment}
\vspace{-4pt}
\input{figures/model_fig}

\subsection{Experimental details}

\noindent \textbf{Dataset.}
\label{dataset}
We used the dataset~\cite{zarkogianni_konstantia_2022_8301142, covid-19-sensor-informatics-challenge}~provided by the challenge organizers for network training and testing.
While both deep breath signals and cough signals were available, we only employed labeled cough signals for our task.
As we are unable to assess the performance on the test set (which is a blind dataset), we allocated 8 \% of the provided dataset to serve as the validation set.

\noindent \textbf{Feature extractor fine-tuning details.}
\label{sec:finetune_config}
We conducted fine-tuning of the Wav2vec 2.0 model over 2000 epochs, employing a batch size of 16 and using the cough sounds dataset.
We employed the Adam optimizer with an initial learning rate of 0.00003 and incorporated a learning scheduling scheme based on polynomial decay.
Upon completing the fine-tuning of the feature extractor, no further gradient updates are applied to the feature extractor.

\noindent \textbf{C2C implementation details.} 
We trained the E-T encoder and classifier over 3900 epochs, with a batch size of 32.
We also employed the Adam optimizer with an initial learning rate of 0.0003 and the  CosineAnnealingWarmUpRestarts~\cite{loshchilov2017sgdr} learning scheduler.

\noindent \textbf{Data augmentation.} 
We applied random shift augmentation on the training dataset~\cite{defossez20_interspeech}, shifting of audio either to the left or to the right by a random number of seconds.
Specifically, we randomly shift 1-second intervals within 4-second audio segments.
We also applied a feature masking augmentation for the feature output, which resembles the technique used in SpecAugment~\cite{park19e_interspeech}.

\noindent \textbf{Evaluation protocols.}
We evaluated the performance of COVID diagnosis using the Area under the ROC curve (ROC-AUC).
While ROC is a probability curve at various threshold settings, ROC-AUC quantifies the degree of separability in the classification results.

\subsection{Experimental results}
\vspace{-2pt}
As analysis on the actual test set was not feasible, all evaluations and analyses were performed on the validation set as previously described in Section~\ref{dataset}.

\noindent \textbf{Experiments with different input signals.}
Fig.~\ref{fig:fig} illustrates the network flow using various input signals, including Cough only, deep breath only, and cough and breath signals.
We employed identical network architectures for all the experiments. 
Table~\ref{tab:various_input} shows that the proposed C2C network achieved an ROC-AUC score of 0.7810.
The comparison results between the two models, D2C and C2C, demonstrate that the inclusion of deep breath signals resulted in a performance degradation.
Furthermore, we observed that the learnable parameter $\alpha$ converged to 1, used in B2C, which also indicates that the inclusion of deep breath input features did not enhance the network performance.

\noindent \textbf{Ablation studies.}
Table~\ref{tab:ablation} demonstrates the individual effectiveness of each proposed process (pre-processing, feature extractor, data augmentation).
The model's training failed to make any progress when pre-processing was omitted, resulting in a ROC-AUC score of 0.5.
The overall results demonstrate the effective contribution of each proposed process to the performance of COVID-19 diagnosis.
\vspace{-3pt}

%% file: figures/model_fig.tex
\begin{figure}[!t]
    \centering
    \includegraphics[width=.9\linewidth]{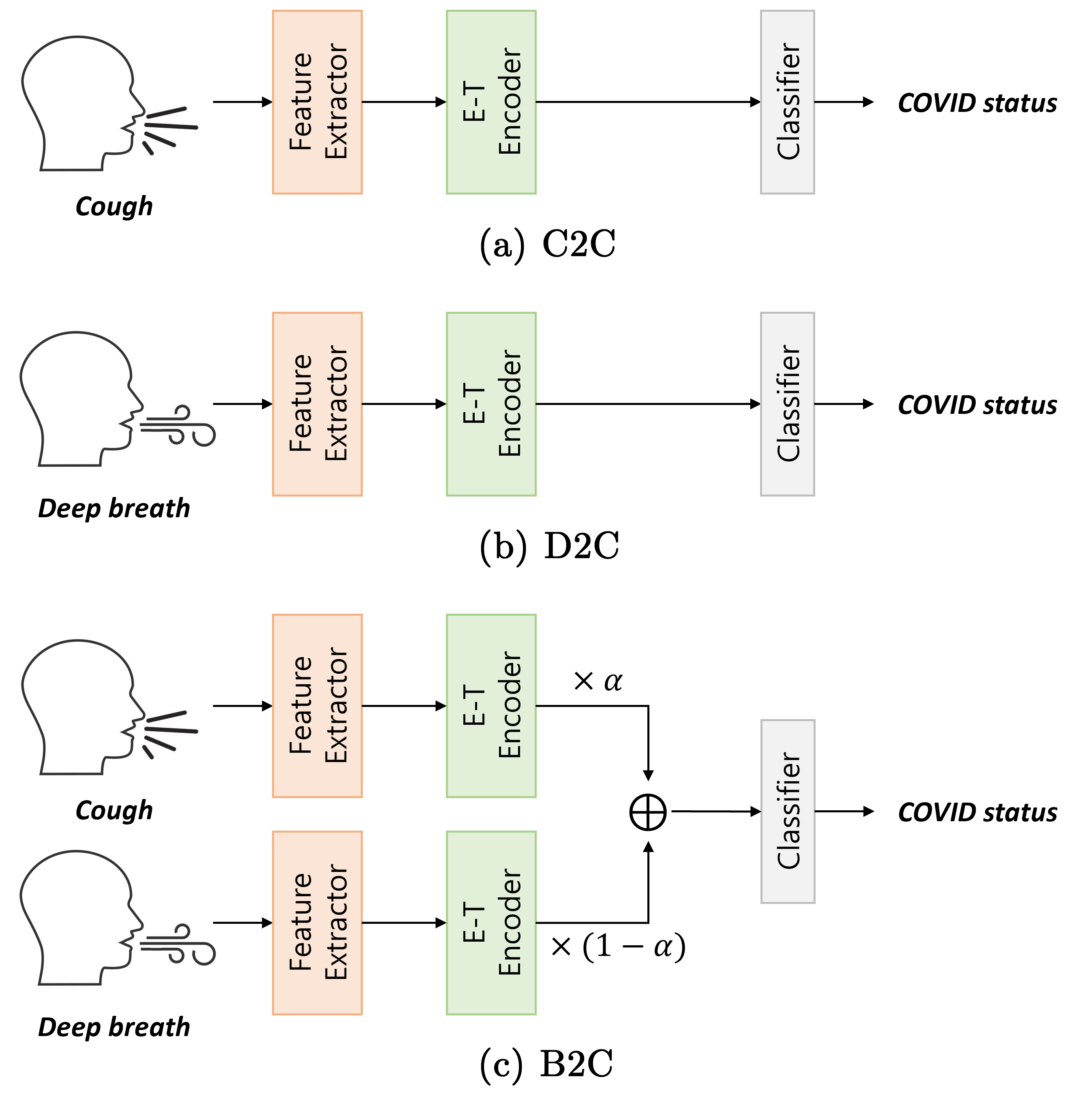}
    \vspace{-8pt}
    \caption{Illustration of the models with different input signals: C2C (Cough to COVID diagnosis), D2C (Deep breath to COVID diagnosis), and B2C (Cough and deep breath to COVID diagnosis). $\alpha$ denotes a learnable parameter.}
    \label{fig:fig}
\vspace{-15pt}
\end{figure}

%% file: 5_Conclusions.tex
\section{Conclusion}
\label{sec:conclusion}
\vspace{-3pt}
\input{tables/various_input}
\input{tables/ablation_study}
This report provides an overview of our system, \shortname, for the BHI 2023 Data Competition: Sensor challenge, designed for COVID-19 diagnosis using patient cough recordings. 
We have observed that adopting data pre-processing, and data augmentation with a powerful feature extractor resulted in a significant improvement in performance.
Experimental results confirm the high potential of \shortname~ in achieving accurate COVID-19 diagnosis.

\vspace{-6pt}

%% file: tables/various_input.tex
\begin{table}[!t]
\centering
\caption{Performances on different input signals}
\label{tab:various_input}
\vspace{-2pt}
\begin{tabular}{l||c}
\toprule
Model                           & ROC-AUC ↑            \\ \midrule
\textbf{C2C (Proposed)}         & \textbf{0.7810}   \\ 
D2C                             & 0.6749            \\  
B2C                             & 0.7310            \\ \bottomrule
\end{tabular}
\vspace{-3pt}
\end{table}


%% file: tables/ablation_study.tex
\begin{table}[!t]
\centering
\caption{Ablation study for each contributing module}
\label{tab:ablation}
\vspace{-2pt}
\resizebox{\columnwidth}{!}
{
\begin{tabular}{l||c|c}
\toprule
Model                           & ROC-AUC ↑          & Performance variation (\%) \\ \midrule
\textbf{C2C (Proposed)}         & \textbf{0.7810}   & -                          \\ 
w/o pre-processing              & 0.5000            & -35.98                      \\ 
w/o feature extractor           & 0.6277            & -19.63                      \\ 
w/o data augmentation          & 0.6729            & -13.84                      \\ \bottomrule
\end{tabular}
}
\vspace{-15pt}
\end{table}